\documentclass[floatfix,aps,rsi,groupedaddress,twocolumn]{revtex4-1}
\usepackage{graphicx}
\usepackage{amsmath}
\usepackage{pdfpages}
\usepackage{url}

\begin{document}

\begin{abstract}
	A chopper wheel construct is used to shorten the duration of a molecular beam to \mbox{13 $\mu$s}. Molecular beams seeded with NO or with Br$_2$ and an initial pulse width of $\geq 200$ $\mu$s were passed through a spinning chopper wheel,  which was driven by a brushless DC {\it in vacuo} motor at a range of speeds, from \mbox{3,000 rpm} to \mbox{80,000 rpm}. The resulting duration of the molecular-beam pulses measured at the laser detection volume ranged from \mbox{80 $\mu$s} to \mbox{13 $\mu$s}, and was the same for both NO and Br$_2$.
%, measured using a Savitzky-Golay fitting method.%
% The duration  depends primarily on chopper rotational speed, regardless of the seed molecule used, and is 
The duration is consistent with a simple analytical  model, and the minimum pulse width measured is limited by the spreading of the beam between the chopper and the detection point as a consequence of the longitudinal velocity distribution of the beam.
The setup adopted here effectively eliminates buildup of background gas without the use of a differential pumping stage, and a clean narrow pulse is obtained with low rotational temperature.  
\end{abstract}

\title{A chopper system for shortening the duration of pulsed supersonic beams seeded with NO or  Br$_2$ down to \mbox{13 $\mu$s}}
\author{Jessica Lam, Chris Rennick, Tim Softley}
\affiliation{Department of Chemistry, University of Oxford, Chemistry Research Laboratory,
12 Mansfield Road, Oxford, OX1 3TA, United Kingdom}
\maketitle

\section{Introduction}

Slow, velocity-controlled molecular beams have become an attractive tool in the study of molecular dynamics, the measurement of fundamental constants and in conducting high resolution spectroscopy \cite{ColdMolApplicationsSchnellMeijer, TamingMolBeamsMeijer, HindsEDMviaStarkDecel, ColdMolForHighResSpectrMeijer, ColdMolForHighResFineStructureConstJunYe}. Many advances have been made to produce slow beams such as using a high-pressure seeded expansion \cite{HighPressureJetExpansion}, pre-cooling from a buffer gas source \cite{BufferGasASBEAMPbOandNaDemille,BufferGasASBEAMO2Doyle}, or by passing the gas through a cryo-cooled valve nozzle \cite{CryocoolValveRSI,CryocoolValvePRL} or a room-temperature nozzle spinning in a counter-rotating configuration \cite{CounterRotatingNozzleMudrich, CounterRotatingNozzleHerschbach, CounterRotatingNozzleSheffieldHerschbach}. Molecules in seeded supersonic beams can also be decelerated by undergoing inelastic 
 collisions in a crossed beam set up to remove kinetic energy \cite{ChandlerBilliardNOAr}, 
or by using the interaction of the electric or magnetic dipole moments of the molecules with external optical dipole, electric, or magnetic fields \cite{OpticalPotentialSlowBeams, StarkDecelOriginalMeijer, timevaryingStarkNeutralsOldGould, ZeemanFirstMerkt, ZeemanFirstRaizen}.  

Many of these recent advances in the production of cold atomic and molecular beams would benefit from using the shortest gas pulse that is compatible with maintaining a high number density, narrow velocity distribution and cold rotational distribution.
%In particular, adiabatic expansion of the gas at very short pulses, converts nearly all of the energy of the gas into translational motion, with velocity spreads and internal energy states close to zero in the moving frame. 
%This allows the formation of seeded beams containing high density, a cold rotational distribution for high state-selectivity, and a narrow velocity distribution \cite{IntroChopperCampargue, helicalvelselectorcontinuousbeams}.
 Very short pulsed beams produce fewer gas molecules in the vacuum chamber, thereby lowering the gas load on the pumps for the experimental apparatus, and reducing background collisions with the beam; %The reduction of background gas buildup in pulsed valve systems can enable higher signal-to-noise measurements, as well as reduce the number of collisions of the molecular beam with the background gas, %
the low velocity spread and cold internal temperature of the molecules in the beam are thus preserved \cite{IntroMolBeamsScoles, PulsedSourcewIonGaugeOld, highRepPulsedValentiniOld}. 

Short-pulse molecular beam sources are primarily pulsed valves that use modified, fast-acting, resonantly-driven valve components. For example, the Jordan Pulsed Supersonic Valve provides cold molecular beams of \mbox{55 $\mu$s} duration during choked flow and \mbox{20 $\mu$s} during low flow \cite{JordanValve}. This valve operates on the magnetic beam repulsion principle by passing opposing currents through two parallel-beam conductors, one of which is a thin metal strip that blocks the gas outlet. The currents produce a magnetic force which moves the strip and thus opens the valve \cite{Dimov}. Another fast-acting pulsed valve was recently constructed by Janssen and colleagues; 
it uses a cantilever piezo to produce molecular beam pulses of \mbox{7 $\mu$s} duration at high repetition rates of \mbox{5 kHz} \cite{JanssenPiezoValve}. The Even-Lavie pulsed valve uses light, miniaturized components to enable operations at high backing pressures of up to \mbox{100 bar}, resulting in a high intensity molecular beam with a duration of \mbox{8 $\mu$s} and with very cold rotational temperatures for entrained molecules of around \mbox{0.5 K} \cite{EvenValveA, EvenValveB}. Most recently, the Nijmegen Pulsed Valve was constructed, by pulsing a current through an aluminum strip located within a magnetic field. The Lorentz force causes the strip to break its seal to the nozzle in a periodic fashion, resulting in molecular beams with durations as low as \mbox{20 $\mu$s} and a rotational temperature of around \mbox{1 K} \cite{Nijmegan}.

These beam sources are typically constructed from many small individual components. The high surface area to volume ratio ratio of these components makes them susceptible to corrosion from redox reactions with chemically reactive seed molecules, such as halogen species \cite{AluminiumCorrosion, CorrosionBrCraigAnderson,CorrosionBrDavis,CorrosionBrUhlig}. Thus, the use of a valve that contains a few robust components, while still realizing a full expansion of the molecular beam, is advantageous to experiments using reactive seed molecules.

The commercially available General Valve Series 9 valve system has a relatively simple design, and the components are robust against corrosion. The main body is %The General Valve is a simple design, %
constructed with corrosion-resistant stainless steel, and a solenoid is used to drive a Teflon cylinder with a conical tip, which breaks the seal to the valve nozzle, producing a fully-expanded molecular beam \cite{GeneralValve}. The Teflon cylinder can be easily replaced, and the buildup of halogen-induced oxidation on the metal components can be easily removed and sanded off without changing the shape of the valve geometry. Thus, the valve is fully compatible with halogen-seeded molecular beams, and residue from the oxidation of the halogen molecules on the valve can be easily washed off with deionized water or acetone. Our current work uses a Series 9 valve to generate a molecular beam of Br$_2$ seeded in Ar.
In these experiments, near-threshold photolysis of Br$_2$ produces cold Br atoms in a so-called `Photostop' experiment, and the atoms are contained in a permanent magnetic trap on the molecular beam axis for up to \mbox{99 ms} \cite{DohertyPhotostop, RennickLamPRL}. A disadvantage of using this type of valve is that the beam pulse duration is typically  on the order of a hundreds of microseconds to a few milliseconds, so the high density of carrier-gas (Ar) atoms in the tail of the pulsed  beam (typically 10$^{13}$ cm$^{-3}$ at the center of the trap) is the dominant cause of trap loss of Br atoms \cite{LamPRA}. In order to accumulate density of Br atoms in the trap from successive pulses of the beam, we need to minimize the trap loss in the \mbox{100 ms} time interval between the successive photolysis pulses. Thus, we need to reduce the pulse-duration of the molecular beam, both to minimize collisions with the tail of the beam, and also to reduce the background pressure in the chamber.

An alternative method of producing short gas pulses, in conjunction with a valve that initially generates long pulses, is to mechanically block the majority of the original molecular beam pulse, and only open the beam path in  a narrow time window to let a thin slice of the pulse width through to the experimental chamber. Schwarz \textit{et al.} constructed a molecular beam chopper using an electromagnet to resonantly drive a magnetic field across a contact pin. This oscillation resulted in the blocking and unblocking of the molecular beam as it passed by the contact pin, with the varying density of the beam closely matching that of a sinusoidal function \cite{SchwarzChopperSinosoidal}. A more common setup is the use of a spinning chopper wheel \cite{GeneralValve, CrossCorrelationTOFMolBeamScattering1979,MagneticCrossCorrelationChopper,microchopperneutralbeams,SmallNeutronBeamChopper,TOFmeasurementsWChopper1977RSI,UHVChopper1980}. The molecular beam is directed perpendicularly to a spinning disk, into which slots are cut. The rotation of the disk is synchronized to the gas pulse so that only a portion is allowed through the slot, and the rest of the beam is blocked. Abad \textit{et al.} also constructed a chopper set up by rotating a copper disk at  \mbox{11,280 rpm} to produce molecular beam pulse durations of around \mbox{23 $\mu$s}, in order to measure the velocity distributions and translational temperatures of the molecular beam from a Series 9 valve \cite{GeneralValve}.

However, two disadvantages of using a chopper wheel to shorten molecular beams have prevented this method from dominating over modified pulsed valves. First, the relatively low mechanical speed of rotation used for chopper wheels to date limits the shortest gas pulse attainable. Increasing the disk radius of the chopper increases the tangential velocity, and hence reduces the pulse width for a given slot dimension. But a larger disk also increases the mechanical and heat load on the system, and the radius is typically limited by the geometry of the vacuum chamber and the ability to balance the disk onto the motor at high revolution speeds. Second, the large portion of the molecular beam that does not pass through the chopper slit may contribute to the  background pressure in the vacuum chamber. In some cases  this gas buildup has been mitigated by placing the chopper wheel along the path of a set of skimmed differentially pumped stages \cite{rotatingchopperskimmer}.

In this paper, we demonstrate the use of a new high-speed chopper wheel construct to produce a \mbox{13 $\mu$s} gas pulse. We furthermore show that this technique can operate with a simplified differential pumping stage. We have applied this technique to shorten beams seeded with corrosive NO and Br$_2$ molecules, for which other short-gas-pulse sources may have limited durability.

\section{Experiment}

\begin{figure}[htb]
\includegraphics{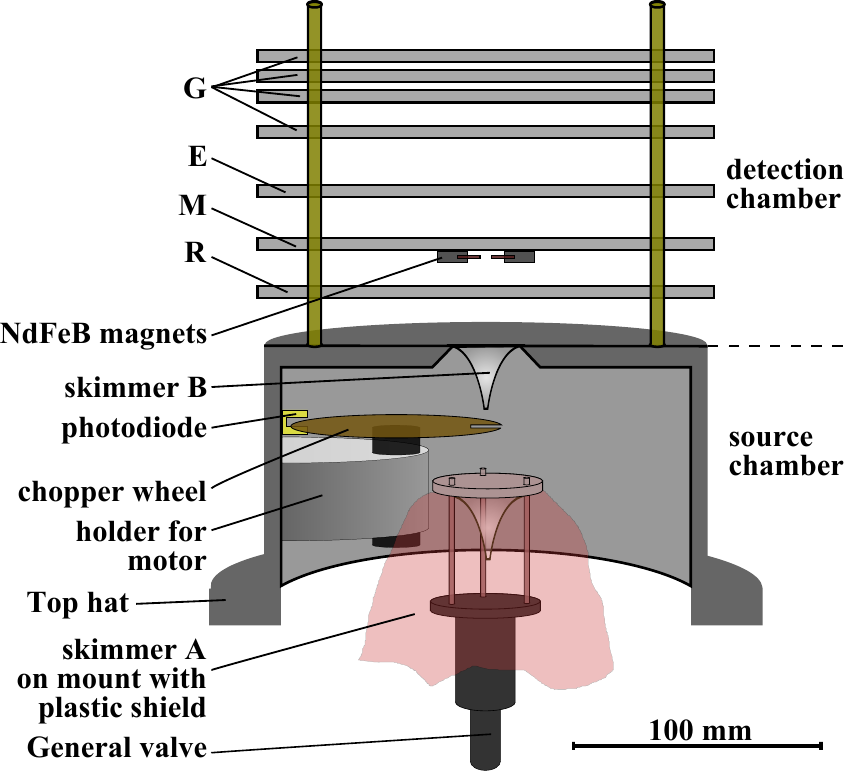}
\caption{Molecular beam and ion detection section of our experimental setup (approximately to scale). The dotted line (right) shows the separation between the source chamber (bottom) and the detection chamber (top).}
\label{fig:expsetup}
\end{figure}

\begin{figure}[htb]
\includegraphics{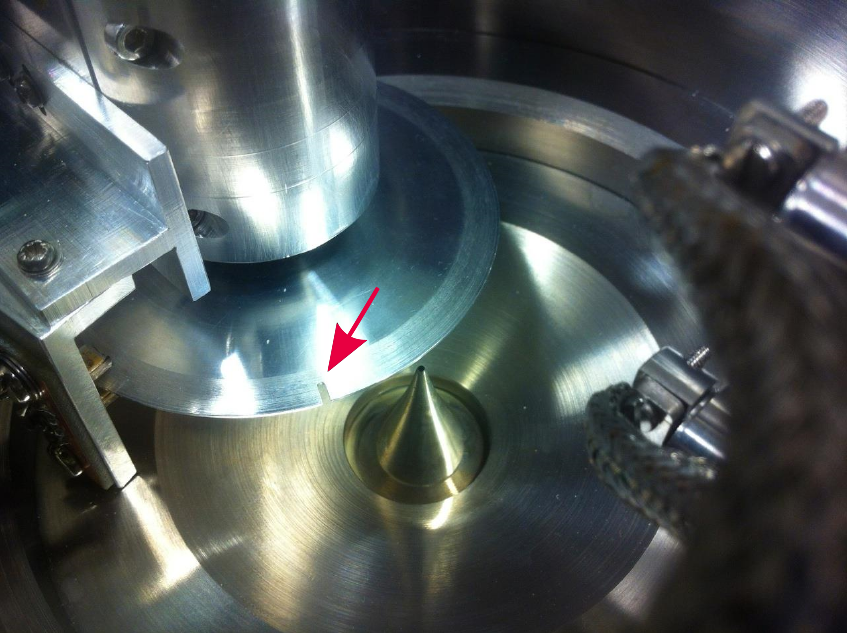}
\caption{Photograph of chopper setup in the source chamber. The chopper slit is seen as the indent in the chopper wheel, indicated by the red arrow.}
\label{fig:expsetup2}
\end{figure}

A schematic of the experimental setup is shown in figure \ref{fig:expsetup}.  The pulsed valve (General Valve, Series 9) generates a molecular beam, which passes through two skimmers separated by the chopper wheel and passes through skimmer B into the differentially-pumped detection chamber.
The chopper, its motor, and the pulsed valve are housed within  the stainless steel `top hat', located at the end of the source chamber. The source chamber is pumped by a \mbox{700 ls$^{-1}$} diffusion pump, while the detection chamber is pumped by a \mbox{350 ls$^{-1}$} turbo pump.
%B and source chamber are situated below a turbo-pumped second chamber, called the detection chamber, which houses the laser detection volume and electrostatic plates. %The source and detection chambers are separated by skimmer-B, as shown in the figure, and maintain a base pressure of \mbox{3$\times$10$^{-7}$ mbar}.
The chopper wheel, shown in Figs. \ref{fig:expsetup} and \ref{fig:expsetup2}, is a \mbox{70-mm} diameter aluminum (7075 Aircraft alloy) disk with two \mbox{1 mm}-wide slits cut \mbox{3 mm} at 180$^\circ$ into the radius of the disk. The disk is thinned at the edge to less than \mbox{1 mm}, and directly coupled to the shaft of a \mbox{25 mm} diameter brushless DC motor (Maxon Motor model EC25) and balanced at up to \mbox{42,000 rpm} in ambient conditions. As shown in Fig. \ref{fig:expsetup} the chopper wheel is placed between skimmer-A and \mbox{skimmer-B} with the slitted edge aligned with the center of the molecular beam.
% as shown in figure \ref{fig:expsetup2}.

The chopper is driven by a home-built controller which generates the AC wave forms used to drive the brushless motor at the desired speed, ranging from \mbox{3,000 rpm} to \mbox{80,000 rpm}. The motor  is cooled by contact with the wall of the vacuum chamber, and remains below \mbox{$80\,^{\circ}{\rm C}$} at speeds up to \mbox{80,000 rpm}.
%The high speed of the chopper was made possible due to the following characteristics of the motor. First,%
 This brushless DC motor is highly suitable for this application and its characteristics make possible the high speed of the chopper. It is compatible with a vacuum environment due to it having fewer components than other types of motor; it has  a high torque output per power input due to the low friction around its bearings; it has low electromagnetic interference; it has an ability to operate without requiring the use of additional cooling agents such as airflow or water \cite{BrushlesDCEDNNetwork, BrushlessDCMaxon, BrushlessDCMicrochip}; finally, the motor uses ball bearings which are vacuum compatible and minimizes the buildup of heat around the rotor responsible for limiting the speed of the motor.

Skimmer-A (orifice diameter \mbox{1 mm})  is rigidly mounted to the General valve, as shown in figure \ref{fig:expsetup}.
This rigid mounting mechanically aligns \mbox{skimmer-A} to the valve orifice at a distance of \mbox{3 cm} from the skimmer tip. The mount is surrounded by a shroud made of plastic polyethylene terephthalate (PET) foil, which forms an effective differential pumping stage. This setup enables only the central part (\mbox{1 mm} diameter) of the molecular beam, passing through the  first skimmer, to reach the region of the chopper wheel. The rest of the beam traveling off-axis is directed by the PET foil to the diffusion pump, which has a pumping speed sufficient to allow the background gas to be pumped away before further buildup from the next gas pulse. Any gas escaping around the edge of the shroud  has a long enough diffusion time that it does not interfere with the fast  on-axis  beam. The molecular beam pulse shortened by the chopper wheel subsequently passes through a \mbox{0.5 mm} skimmer (B), which blocks further background gas scattered by the chopper wheel from reaching the detection chamber. 
%The chopped gas pulse then travels to a second differentially-pumped chamber where it is intersected by a nanosecond probe laser. This laser ionizes the gas in a probe volume, and these ions are detected by a conventional Wiley-McLaren time-of-flight and microchannel plate. The delay between the lasers and gas pulse is scanned, and the ion signal at each delay is proportional to the beam density.

% \begin{figure}[htb]
% \includegraphics{2ndskimmerb}
% \caption{Photo of the second skimmer attached to the General valve, side view.}
% \label{fig:expsetup3}
% \end{figure}

In the detection chamber, the beam passes through a series of electrostatic plates, labeled G (ground), E (extractor), M (magnet), and R (repeller), designed for time-of-flight detection of ions, produced by resonance-enhanced multiphoton ionization (REMPI) at a position close to plate M. Two NdFeB magnets are mounted on  plate M, and can be used to confine low-velocity paramagnetic species  in the Photostop experiment (but are not used as such in the experiments reported here; see \cite{RennickLamPRL} for further details). Two laser beams pass between the two magnets parallel to plate M in counter-propagating directions. Laser 1, a \mbox{10 Hz}-pulsed dye laser beam (\mbox{5 ns} pulse length) is used to photodissociate molecules in the beam (Br$_2$/Br experiments only).  Laser 2, a \mbox{10 Hz} frequency doubled dye laser, is the probe laser and is used for REMPI of either the NO molecules or the Br fragments.

\begin{figure}[htb]
\includegraphics{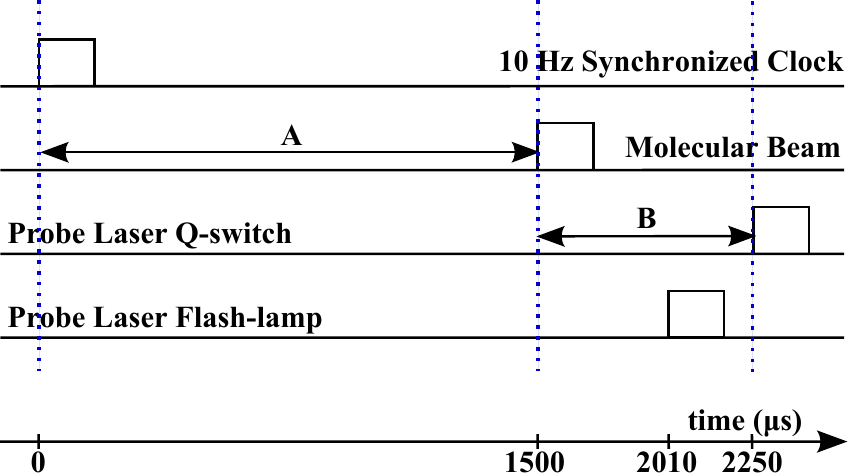}
\caption{Diagram of timing setup in the chopper controller and delay generator. 
%The time delays are indicated by \textbf{A} and \textbf{B}. 
\textbf{A} is the time between the \mbox{10 Hz} clock signal generated by the chopper and the trigger for the pulsed valve actuation, and \textbf{B} is the delay between the pulsed valve and the  dissociation/probe laser Q-switch triggers. The widths of the trigger pulses are on the order of \mbox{10 $\mu$s}.
}
\label{fig:timingpulse}
\end{figure}

The chopper generates the clock signal used to trigger the pulsed valve and the probe laser(s). The chopper itself serves as the base clock, thereby eliminating the need to synchronize the chopper phase to an external clock. As the edge of one of the two chopper slits passes between the photodiode (shown in figure \ref{fig:expsetup2}) and a LED light, a trigger pulse is passed back to the chopper controller through a pair of  vacuum feedthroughs (Allectra GmbH). This trigger pulse train is corrected to a \mbox{10 $\mu$s}-wide pulse, independent of rotation frequency, and a gate allows only the first pulse in every \mbox{100 ms} through to the output. This output is therefore synchronized to the chopper and nominally at a \mbox{10 Hz} frequency demanded by the pulsed laser system. The actual frequency is allowed to vary between \mbox{9.5 Hz} and \mbox{10.5 Hz} to maintain the synchronization, with negligible impact on the laser power. This pulse triggers the input of a delay generator (Quantum Composer 9500 Plus Series), which in turn, triggers the pulsed valve (after delay A) and the probe lasers after  a further delay B, as shown in figure \ref{fig:timingpulse}.

\section{Analytical model of pulse width reduction}

The terminal velocity of a molecular beam, $v_\infty$, is given by \cite{IntroMolBeamsScoles}

\begin{equation}
{v}_{\infty}=\sqrt{\frac{\bar{\gamma}}{\bar{\gamma}-1}\frac{2k_\text{B}\mathit{T}}{\mathit{m_\text{eff}}}}\label{Eqmuinf}
\end{equation}

\noindent
where $\bar{\gamma}$ is the mean heat capacity ratio of the mixture, calculated to be $1.67$ for a weighted average of 8\% Br in Ar and $1.63$ for 8\% NO in Ar; $\mathit{k_{B}}$ is the Boltzmann constant, $\mathit{T}$ is the initial temperature of the gas in the reservoir ($\mbox{298 K}$), and $\mathit{m_\text{eff}}$ is the weighted average mass of the mixture \cite{IntroMolBeamsScoles}. The standard deviation of velocities around $v_\infty$ in the moving frame is related to the translational temperature of Ar in the moving frame, $\mathit{T_\text{Ar}}$.
\begin{equation}
v_\text{s}=\sqrt{\frac{2\mathit{k_\text{B}}\mathit{T_\text{Ar}}}           {\mathit{m_\text{Ar}}}   }\label{vs}
\end{equation}

\noindent
The packet of molecules that passes through the chopper has a length $\omega_0$ (immediately after the chopper wheel) that is inversely proportional to the rotational frequency of the wheel
and is given by
\begin{equation}
w_\text{0}=v_\infty \frac{d_\text{slit}}{2\pi r S} \label{eq:w0}
\end{equation}
\noindent
where $d_\text{slit}$ is the width of the slit tangential to the edge of the chopper (equal to \mbox{1 mm}) $r$ is the radius of the chopper wheel, and $S$ is the rotational frequency of the chopper in revolutions per second (note that in the results reported below, the rotation frequency of the chopper wheel is quoted in revolutions per minute (rpm)). 
%\begin{equation}
%t=\frac{d}{v_\infty}  \label{t}.
%\end{equation}
%\noindent
The moving-frame temperature causes longitudinal expansion of the packet during this flight time, with the spatial spreading given by
\begin{equation}
w_\text{s}= v_{\text{s}} t = {v}_{\text{s}} \frac{d}{v_\infty} \label{eq:ws}
\end{equation}
\noindent
where $d$ is the distance between the chopper and intersection with the detection laser, and $t$ is the mean time taken for the packet to travel that distance.
Using equations (\ref{eq:w0}) and (\ref{eq:ws}), the length of the detected packet at the laser detection volume, $w_\text{beam}$, is  obtained from the convolution of the length of the initially chopped packet with  that due to its longitudinal velocity distribution $v_\text{s}$. A Gaussian packet and velocity distribution is assumed to give
\begin{equation}
w_\text{beam}^2=w_\text{0}^{2}+w_\text{s}^{2} = (v_\infty \frac{d_\text{slit}}{2\pi r S})^{2}+(v_{\text{s}} \frac{d}{v_\infty})^{2}\label{eq:wbeam}
\end{equation}
\noindent
At low rotational frequency of the chopper, the spatial packet size of the molecular beam is determined  by  the first term in equation \ref{eq:wbeam}, while at high frequency it is determined  by the spread of velocities around the moving frame, $v_\text{s}$, as indicated by the second term.

\section{Results}
\subsection{Measurements with molecular beam of NO}

The chopper setup was first tested with a molecular beam composed of 8\% NO seeded in $\mbox{3 bar}$ Ar backing gas. NO was chosen due to its strong, well-resolved, rotational lines in the REMPI spectrum (enabling ease of calibrating the rotational temperature of the molecular beam from the intensities) and the convenient wavelength of the probe laser in our setup. The photodissociation laser was not used for these experiments. The NO is detected %
%in the center of the magnetic plate (labeled \textbf{M} in figure \ref{fig:expsetup})%
 by (1+1) REMPI via the $A^{2}\Sigma^{+} \leftarrow X^{2}\Pi_{\frac{1}{2},\frac{3}{2}}$ transition at a wavelength around $\mbox{226.2021 nm}$ \cite{NOspectra, NO11REMPIZare}. The fundamental wavelength was generated from a pulsed dye laser (Sirah Continuum, $\mbox{Coumarine 503}$ dye) and doubled with a BBO crystal. The NO$^+$ cations produced by REMPI close to plate M are guided towards the MCP detector (absent from figure \ref{fig:expsetup}), by the field applied across plates R, M and E, using a Wiley-McLaren configuration \cite{WileyMcLaurenTOFMassSpec}.

The rotational temperature of NO in the molecular beam was determined by recording the REMPI signal of NO while scanning the detection wavelength $\mbox{226.000 nm}$ to $\mbox{226.400 nm}$. The rotational line intensities were fitted to both a Boltzmann plot, and also to the full simulation of the spectrum using the simulation program PGOPHER %used to calculate the rotation, vibrational and electronic spectra of molecules
\cite{PGOPHER}. Using both methods, the distance between the chopper and the pulsed valve was set such that a lowest measured value of the rotational temperature of the NO in this molecular beam was obtained at \mbox{7 K}. The low rotational temperature indicates that there was little or no buildup of background gas behind the chopper wheel to interfere with the molecular beam passing through it.

\begin{figure}[htb]
\includegraphics{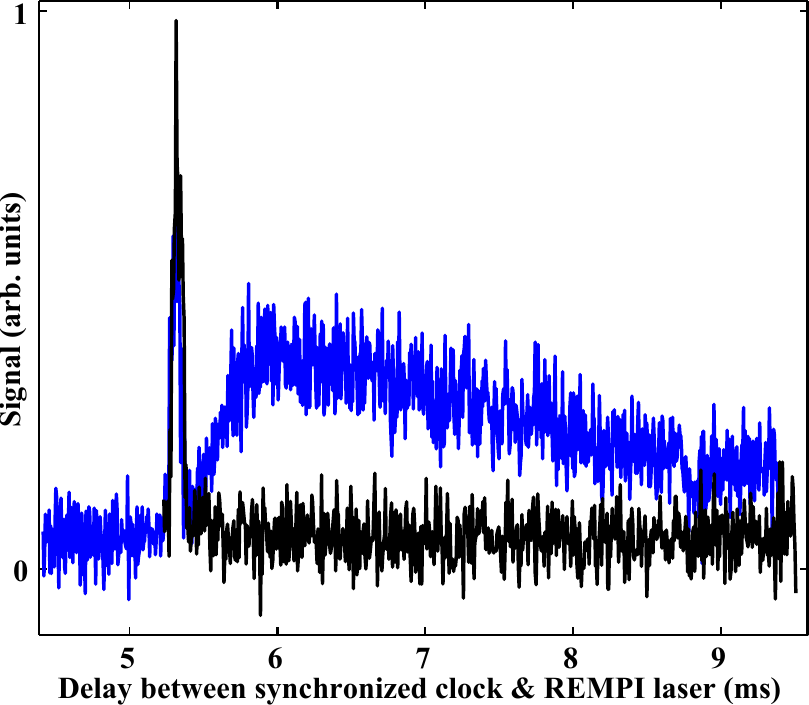}
\caption{Comparison of signal of NO with a chopper running at a revolution speed of \mbox{3,000 rpm}, with (black) and without (blue) the plastic shield.}
\label{fig:plasticshield} 
\end{figure}

The NO experiment was  used to test the feasibility of the effective differential pumping stage using the PET foil. Figure \ref{fig:plasticshield} shows the signal of NO with the probe laser fixed at a wavelength of $\mbox{226.2021 nm}$ 
(corresponding to the position of the overlapping  Q$_{21}$\ (0.5)  R$_{11} (0.5)$  lines)
 and taken with a chopper speed of \mbox{3,000 rpm}, with and without this plastic shroud. 
The delay between the synchronized clock signal from the chopper and the probe laser is scanned to record the temporal profile of the beam passing through.
In both cases, a clear  sharp signal of NO indicates that a molecular beam pulse of the same duration passes through the chopper wheel, as shown at a delay of around \mbox{5.3 ms} in figure \ref{fig:plasticshield}. However, in the case without using the PET foil, a broad intense signal of NO is observed at longer time delays for which we would expect  the chopper to have shut off the path of the molecular beam. The broad signal results from gas that does not pass through the first skimmer but diffuses to the detection region.
The broad signal is eliminated when using  the PET foil; the foil acts as a differential pumping stage, and therefore allows observation of a clean sharp peak in the time of flight.

\begin{figure}[htb]
\includegraphics{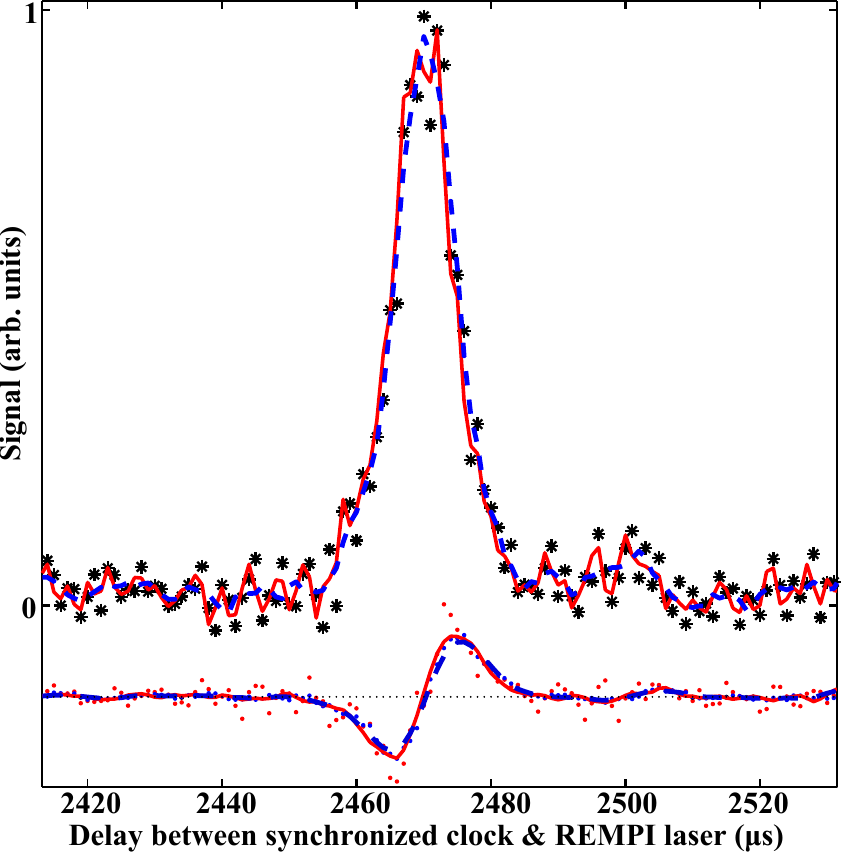}
\caption{A normalized profile of the molecular beam with a chopper running at a revolution speed of \mbox{70,000 rpm}. The upper trace shows the signal intensity, and the lower shows the negative of its first derivative used to define the pulse duration. The series of fittings here is a step-by-step process using a Savitzky-Golay fitting method.}
\label{fig:GolayFit}
\end{figure}

For a range of revolution speeds, the temporal profile of the NO REMPI signal was recorded %for a series of time delays between the phase-locked clock and the REMPI laser, represented by \textbf{A+B} in figure \ref{fig:timingpulse}. This delay accounts for the time for the chopper slit to move from the optical interrupter into the gas beam and the flight time of the gas packet from the pulsed valve to the chopper slit. %
with delay \textbf{A}  set to a constant optimized value of \mbox{750 $\mu$s} to allow the largest gas density 
%through the slit. Delay \textbf{A} is optimized to allow the highest density
 of the original molecular beam packet to reach the chopper slit.
Figure \ref{fig:GolayFit} shows an example of such a profile at \mbox{70,000 rpm}.
A Savitzky-Golay method was used to smooth  the raw data (black stars) \cite{GolayArticle,GolayBook}. This method applies a 2nd order polynomial least-square fit to the data set, as shown in figure \ref{fig:GolayFit}, where the \mbox{solid-red} and \mbox{dashed-blue} curves in the fit above the $y=0$ axis indicate fine and rough fits, respectively, according to the number of data points used in the smoothing algorithm. Next, the first-derivative is taken from these curves, and the Savitzky-Golay method is applied again, as shown by the curves located below the zero $y=0$ axis in figure \ref{fig:GolayFit}.
The beam width is defined as the difference between the delays on the \mbox{x-axis} representing the peak and zenith of this final, smoothed, first derivative. The beam width for the profile in figure \ref{fig:GolayFit} is determined to be \mbox{8 $\mu$s} and \mbox{11 $\mu$s}, from the fine and rough fits, respectively.

\begin{figure}[htb]
\includegraphics{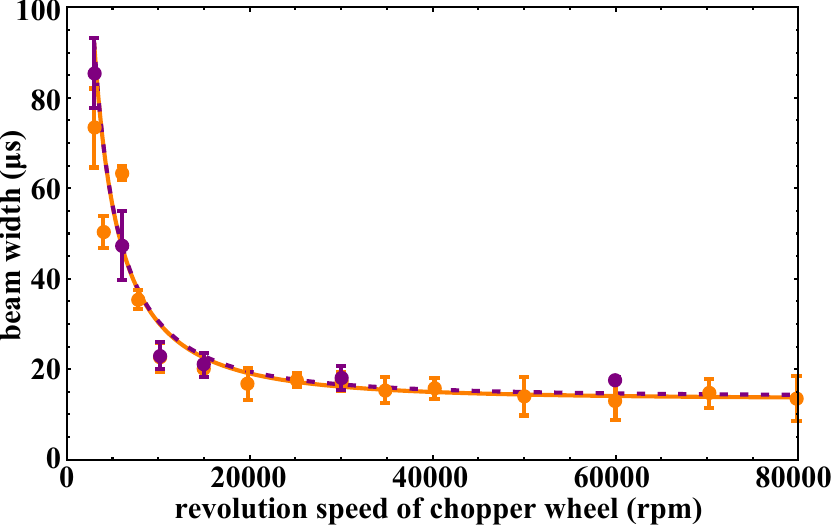}
\caption{Plot of beam widths of NO (orange) and Br (purple) as a function of chopper speed. Each data point represents an average and first standard deviation of three sets of beam-width measurements. The theoretical model, $w_\text{beam}$, described by equation \ref{eq:wbeam} is represented by the solid-orange and dashed-purple curves for NO and Br$_2$  respectively.}
\label{fig:Golay}
\end{figure}

A series of these beam widths was determined for NO over a range of revolution speeds of the chopper wheel, from \mbox{3,000 rpm} to \mbox{79,800 rpm}, and the values are plotted as the orange points in figure \ref{fig:Golay}. 
%Experimental measurements with Br$_2$ are described in the next section. 
Each data point is an average of three beam-profile measurements, each fitted using a rough fit (dashed-blue in figure \ref{fig:GolayFit}) to provide an upper limit to the calculated beam widths. The error bars show the variation to one standard deviation of this fluctuation over repeated runs. Thus the value of the lowest averaged beam width of \mbox{13 $\mu$s} for a revolution speed of \mbox{80,000 rpm} is greater than some of the individually measured beam widths, such as that shown in figure \ref{fig:GolayFit}.

The theoretical model value of $w_\text{beam}$, defined by equation \ref{eq:wbeam}, was used to calculate the expected temporal width of  the pulse and is plotted for an NO seeded beam  (solid orange line) in the figure. In both the experimental measurements and theoretical model, beam widths at revolution speeds beyond \mbox{20,000 rpm} converge toward a single value on the order of \mbox{10 $\mu$s}, limited by the longitudinal velocity distribution, $v_\text{s}$, of the molecular beam.

\subsection{Measurements with Molecular Beam of Br$_2$}

To show that the results from the chopper were consistent with other seed molecules, particularly with corrosive halogen species - a key objective of our work -  measurements were also taken with a molecular beam of 8\% Br$_2$ seeded in Ar at a backing pressure of \mbox{3 bar}.
To detect the temporal profile of the Br$_2$ molecules arriving at the laser detection volume, a \mbox{460 nm} Nd:YAG pumped pulsed dye-laser saturates the \mbox{B$^{3}\Pi^{+}_u \leftarrow X^{1}\Sigma^{+}_g$} transition, leading to prompt dissociation to form Br fragments -  for each dissociated Br$_2$, a pair of ground state Br($^{2}\text{P}_{3/2}$) and spin-orbit excited Br($^{2}\text{P}_{1/2}$) fragments are produced. The ground state Br($^{2}\text{P}_{3/2}$) atoms are ionized via ($2+1$) REMPI at $\mbox{226.2021 nm}$ by the probe laser \cite{BrREMPICooper} and directed to the MCP detector. The dissociation laser is fired \mbox{145 ns} prior to the REMPI probe laser, to ensure that the Br atoms are detected at the instant they are dissociated in the probe laser volume. Note that the primary reason for using this two-step detection of Br$_2$ was that our laser system was set up for Br detection in the Photostop experiments \cite{DohertyPhotostop}. 

\begin{figure}[htb]
    \includegraphics{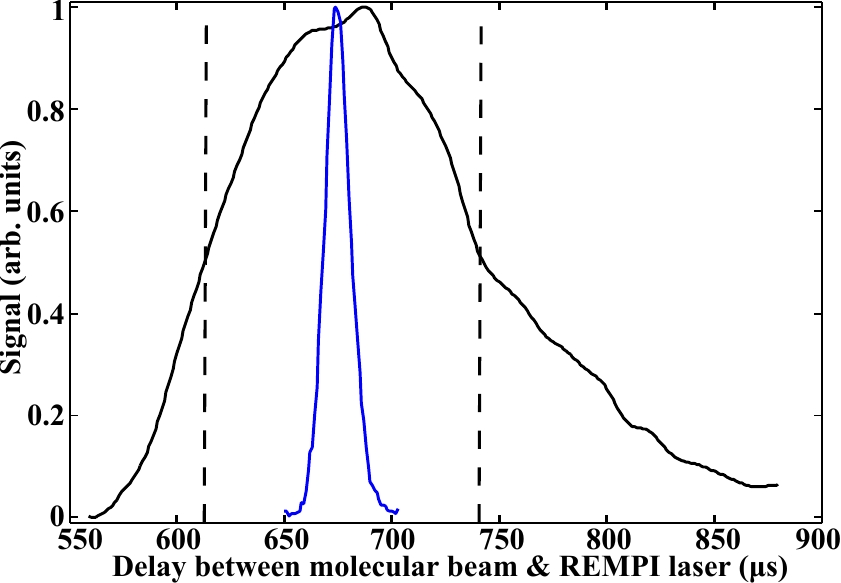}
    \caption{Molecular beam profile of 10\% Br$_2$ seeded in 3 bar Ar without the presence of a chopper (black) and with a chopper running at \mbox{60,000 rpm} (blue). The dotted lines show boundary of the full-width half-maximum (FWHM) molecular beam profile without the chopper.}
\label{fig:molbeamcompare}
\end{figure}

As shown in figure \ref{fig:molbeamcompare}, without the use of a chopper (black line), the duration of the molecular beam has a full-width half-max (dashed lines) of \mbox{130 $\mu$s}, whereas with a chopper running at \mbox{60,000 rpm} (blue) -- around 75\% of its maximum speed -- the duration of the beam is reduced to \mbox{18 $\mu$s}. 
\noindent
In figure \ref{fig:molbeamcompare}, the  molecular beam profile has been smoothed using a Savitzky-Golay algorithm on the experimental data points. Note that although the intensities  of each signal are shown as having the same peak value, the relative normalization is only approximate; the removal or re-installation of the chopper requires substantial changes to the setup and therefore it is difficult to ensure that there is no residual dependence of the  signal intensity on the power  and the alignment of the dissociation and probe lasers in the two sets of measurements. In practice such measurements are separated by days.  Nevertheless it is clear that there is substantial shortening of the molecular beam  pulse.  In figure  \ref{fig:Golay} the measured and calculated pulse widths for the bromine beam are plotted in purple, and these are seen to fit on the same line as the data for NO, verifying the simple analytical model.

%The molecular beam profile without a chopper construct (black) is taken without the timings set by \ref{fig:timingpulse}, as the General Valve in this case is triggered off of an external source, as opposed to a mbox{10-Hz} phase-locked clock from the chopper wheel. For the purpose of comparing the widths of both molecular beam profiles, the \mbox{x-axis} of the profile with the chopper construct (blue) has been moved to overlap with the original source beam (black).

The chopper wheel was designed to ensure that the arrival of the next slot on the chopper wheel will not overlap with the front or tail of the molecular beam. At the highest revolution speed of \mbox{80,000 rpm}, the period between the slits located on opposite sides of the chopper wheel is \mbox{375 $\mu$s}. Thus, even if the very front of the molecular beam were to pass through the chopper slit, the tail of the beam would have been blocked by the chopper wheel before the next slit intersects the molecular beam axis.

It is interesting to note in the context of the Br Photostop experiments, that for the highest chopper wheel rotational speeds, there is a strong correlation between the longitudinal velocity of a Br$_2$ molecule and its arrival time at the laser probe volume - faster molecules in the distribution arrive earlier.
Hence the nanosecond-pulse dissociation laser picks out a subset of the molecular beam pulse with a narrow velocity distribution - much narrower than for the overall pulse.  Thus, in the Photostop experiment the wavelength of the dissociation laser needs to be specified more precisely in order to match the molecular frame photofragment velocity to the lab-frame molecular beam  velocity.  This means that a higher proportion of the molecules that are dissociated will be photostopped, which is beneficial for the trapping experiment.
\section{Conclusion}

In summary, a novel home-built molecular-beam chopper system has been used to shorten the duration of a molecular beam by a factor of ten to \mbox{13 $\mu$s}. Our design demonstrates that a second differential pumping stage is not needed by mounting a second skimmer (A) to the pulsed valve, attached with a plastic shroud to guide the skimmed molecules toward the diffusion pump. With just a few robust components, this method is particular useful for experiments using corrosive species such as halogens as the seed molecules in molecular beams.

The use of a brushless DC motor enables the chopper to run at a range of speeds from \mbox{3,000 rpm} to \mbox{80,000 rpm}, limited only by the cooling of the motor. This range of speeds produces a range of molecular beam widths ranging from \mbox{80 $\mu$s} to \mbox{13 $\mu$s}. The beam widths are repeatable and fall within the same error bars, regardless of whether NO or Br$_2$ is used as the seed molecule.
%A Savitzky-Golay fitting method is shown to be an effective method to
 Measurements of  the beam widths as a function of chopper wheel revolution speed closely match our theoretical model, $w_\text{beam}$, plotted in figure \ref{fig:Golay}.
The ability to use such a simple, yet durable system, offers an opportunity to expand the breadth of chemical systems for which short-pulse molecular beams can be used.

Further work can be done to decrease the pulse duration below \mbox{13 $\mu$s}, by a combination of increasing the revolution speed of the chopper wheel, which requires additional cooling of the motor rotor to maintain its temperature below \mbox{$80\,^{\circ}{\rm C}$}, and by cooling the valve to decrease the spread of velocities around the moving frame of the carrier gas.
%, which requires a greater efficiency in the rotational cooling of the molecular beam as it exits the pulsed valve.

\begin{acknowledgements}
TPS acknowledges the financial support of the EPSRC under grants EP/G00224X and EP/I029109 and the Wiener Anspach Foundation. JL is grateful for support from the NSF Fellowships Program and CJR from the Ramsay Memorial Trust.
\end{acknowledgements}
%\bibliography{ChopperRSIref}

%

\end{document}